# Relative Resolution:

# An Analysis with the Kullback–Leibler Entropy

*Mark Chaimovich[1a], and Aviel Chaimovich[2b]*

[1] *Russian School of Mathematics, North Bethesda, Maryland 20852*
[2] *Department of Chemical and Biological Engineering, Drexel University, Philadelphia, Pennsylvania 19104*

[a] *mark.chaimovich@russianschool.com*
[b] *aviel.chaimovich@drexel.edu*

**Abstract**

A novel type of a multiscale approach, called Relative Resolution (RelRes), can correctly retrieve the behavior of various nonpolar liquids, whilst speeding up molecular simulations by almost an order of magnitude. In this approach in a single system, molecules switch their resolution in terms of their relative separation, with near neighbors interacting via fine-grained potentials yet far neighbors interacting via coarse-grained potentials; notably, these two potentials are analytically parameterized by a multipole approximation. Our current work focuses on analyzing RelRes by relating it with the Kullback–Leibler (KL) Entropy, which is a useful metric for multiscale errors. In particular, we thoroughly examine the exact and approximate versions of this informatic measure for several alkane systems. By analyzing its dependency on the system size, we devise a formula for predicting the exact KL Entropy of an "infinite" system via the computation of the approximate KL Entropy of an "infinitesimal" system. Demonstrating that the KL Entropy can holistically capture many multiscale errors, we settle bounds for the KL Entropy that ensure a sufficient representation of the structural and thermal behavior by the RelRes algorithm. This, in turn, allows the scientific community for readily determining the ideal switching distance for an arbitrary RelRes system.

# 1 Introduction

One of the main challenges of molecular simulations is providing an adequate description of natural phenomena, while maintaining a reasonable computational efficiency. Traditionally, many efforts merely modified the molecular interactions, whilst not altering any aspect of the degrees of freedom[1-4]. Recently, multiscale algorithms linking between detailed fine-grained (FG) and simplified coarse-grained (CG) models became especially popular[5,6]. On a most basic level, these procedures reduce the number of degrees of freedom, by mapping a few FG sites on a single CG site, and in turn, they significantly speed up molecular simulations. However, these strategies usually require complex numerical optimizations from the FG parameters to the CG parameters[7-10], and unfortunately, they still have significant transferability and representability issues in properly describing the natural phenomena of interest[11,12].

Although many schemes have been devised for combining these two molecular models in a single molecular simulation[13-22], especially useful strategies are based on switching between the FG and CG interactions in terms of a spatial variable[23-32]. One such algorithm is called Relative Resolution (RelRes), and its main signature is that it switches between the FG and CG molecular resolutions based on the relative separation: For near neighbors, molecules experience the FG potential, and for far neighbors, molecules experience the CG potential[29,31]. The general approach was initially conceived by Izvekov and Voth, and they applied it on several liquids, most notably on an aqueous solution of a lipid bilayer[29]; Shen and Hu effectively continued applying this variant on other biomolecular systems (it appears that the original formalism was not properly referenced)[30]. Chaimovich et al. then made a couple of important modifications in the preliminary RelRes framework[31,32]. Firstly, instead of switching between the two models in terms of the CG pairwise distance like ref 29, the RelRes Hamiltonian was reformulated so that it switches between the two models in terms of the FG pairwise distance (this practically facilitates the computational implementation of RelRes)[31]. Secondly, rather than performing a numerical parameterization between the FG and CG models as in ref 29 (via "force matching"), an analytical parameterization between the FG and CG models was formally conceived via a multipole series (this alteration naturally removes the requirement for using a computer cluster in this portion of the RelRes scheme)[32]. In turn, it was thoroughly shown for the Lennard-Jones (LJ) potential that RelRes overcomes the transferability and representability issues common with most multiscale algorithms: Across many temperatures and densities, RelRes can replicate structural correlations, thermal properties, etc. of various nonpolar fluids[31,32].

Naturally, each RelRes molecule embodies both FG and CG models, maintaining all coordinates at all times: In the usual recipe, the FG sites are intrinsically represented by atoms, and the CG sites are virtually placed around centers. In such a manner, RelRes preserves all degrees of freedom. So how can this multiscale algorithm enhance computational efficiency? While RelRes does not at all reduce the number of degrees of freedom, it definitively reduces the number of interactions for calculation: As compared with conventional strategies, the short-range interactions between near neighbors are inherently unaltered, yet the long-range interactions between far neighbors are prominently transformed from many FG potentials to few CG potentials; the significant decrease in the number of the latter means that the computer spends less time in calculating them. The fact that RelRes markedly speeds up molecular simulations by almost an order of magnitude was recently proven in LAMMPS[33,34]. In particular, we implemented RelRes in LAMMPS for the LJ potential, and we thoroughly demonstrated that it can correctly describe the static and dynamic behavior of various liquid alkanes, including oligomers and polymers[33]. For all systems that we examined, RelRes enhances the computational efficiency by a factor of 4-5 as compared with conventional strategies. In practical terms for a computer cluster that is evolving thousands of thousands of LJ sites while examining some fundamental non-uniform (multi-component or multi-phase) phenomena, this roughly means that if a typical study may take more than a half year, a RelRes study can likely be less than a couple months.

Given an arbitrary molecule, with a certain mapping between the FG and CG sites, almost all parameters of the RelRes system can be promptly retrieved on a paper sheet rather than on a computer cluster: For example in the case of the LJ potential, the CG energy and length parameters are readily obtained via formulae involving the FG energy and length parameters[31]. However, there remains a single free parameter in the RelRes system that does not ensue the parameterization based on the multipole approximation: That is the "switching distance" at which the molecular resolution changes from the FG interactions to the CG interactions. How is the value for this parameter then chosen? In ref 33, for a selection of alkanes, we performed many molecular simulations at different values of the switching distance, and we examined the capability of RelRes in adequately capturing various structural correlations and thermal properties. We in turn recommended basic rules for our rudimentary alkanes: A switching distance between 0.6 and 0.7 nm (depending on the particular number of FG sites mapped into a CG site), gives sufficient replication in all cases. However, which switching distance shall we use for other systems? Based on ref 33, we can presume some intuitive rules for other alkanes, but how about other organics that involve Coulombic charges? In principle, we must again perform many molecular simulations at different values of the switching distance, and we must examine the capability of RelRes of adequately capturing the static and dynamic behavior of interest. Obviously, this route is extremely tedious.

Fortunately, an informatic measure recently appeared in the multiscale community which can significantly facilitate the process of selecting the switching distance. Here, we call this informatic measure the Kullback–Leibler (KL) Entropy[35]. The KL Entropy has been most often used for multiscale optimization from pure FG systems to pure CG systems[36]. By its mathematical definition, it measures the variation in the probability distributions between two different ensembles. In the context of statistical mechanics, the KL Entropy effectively measures the Hamiltonian fluctuations between two different systems: In this sense, two very useful formulae exist (one exact and one approximate), which reveal this correspondence with the Hamiltonian fluctuations[37,38]. Importantly, it has been shown that the KL Entropy is connected with many "multiscale errors" (i.e., differences in structural correlations, thermal properties, etc. between two different systems)[37,38]. As such, one can ideally develop basic rules for selecting the switching distance for RelRes in various systems: Rather than running many molecular simulations, while exhaustively examining their static and dynamic behavior, one can solely execute a single program which estimates the KL Entropy.

The goal of this article is in evaluating the ability of the KL Entropy in being the guide in selecting the switching distance necessary for a RelRes system. As such, we explore four alkanes which are modeled by the RelRes version of the LJ potential in LAMMPS[33,39]. Foremost in our current work, we examine both the exact and approximate formulae for the KL Entropy. We examine the dependency of both on the system size, interestingly finding that the exact value is a prominent function of the system size, while the approximate value is a negligible function of system size. Based on this, we develop a Taylor-based relation for adequately estimating the exact KL Entropy of "infinite" systems by solely calculating the approximate KL Entropy of "infinitesimal" systems; we importantly find that this empirical formula holds across state space. In turn, we explore the relationship of both the approximate and exact values of the KL Entropy with multiscale errors in structural correlations, thermal properties, etc., establishing values for them that ensure an adequate representation of the system of interest. Overall, this can significantly facilitate the process of selecting the switching distance for a RelRes system.

## 2 Methods

The main goal of our work is to analyze the competence of RelRes via the KL Entropy. Thus, we foremost briefly recap the theoretical foundation of the RelRes framework[31,32], putting a special emphasis on its recent implementation in LAMMPS[33,39]. We then present the KL Entropy[35], specifically in the context of measuring multiscale errors[37,38]. We ultimately describe the setup of the molecular simulations of our alkane systems.

### 2.1 RelRes Fundamentals

Consider a molecular system. Each molecule is composed of several sites; in the current work, the term site typically signifies a "united" atom (e.g., a principal carbon with its adjacent hydrogens). In RelRes, the various sites of a molecule are mapped into several groups (each group is just a collection of sites bonded together); note the special case in which a molecule contains just a single group. In fact, there are two categories of sites in RelRes, "hybrid" sites and "ordinary" sites. A hybrid site embodies both FG and CG potentials; there is just a single such site in a group, and it is typically located around its center. An ordinary site embodies just FG characteristics with no CG features; there can be any number of such sites in a group, and they are typically located at its periphery. In such a manner, the FG and CG models are combined in a single molecular simulation without losing any degrees of freedom. Figure 1 elaborates on this concept, showing two arbitrary groups, which are not necessarily of the same molecule. The groups are labeled with Latin letters $i$ and $j$ (an arbitrary group is occasionally denoted by the index $l$); the sites are labeled with Greek letters $\mu$ and $\nu$ (an arbitrary site is occasionally denoted by the index $\upsilon$). Just like ordinary sites, hybrid sites are naturally labeled by a Greek index $\upsilon$, but because of their uniqueness, we will succinctly label them by a Latin index $l$. Finally, note that a main aspect of a group is its "mapping ratio" $n_l$ (i.e., the number of total sites $v$ mapped onto a group $l$). For clarity, we usually omit this index on $n$.

Such an approach allows switching the resolution of a group from a FG model to a CG model in terms of the relative separation: Usually, far neighbors interact via CG potentials (just of hybrid sites), yet near neighbors interact via FG potentials (also of ordinary sites). We will now define the "switching distance" $r_s$ at which interactions swap between the FG potential and the CG potential; realize that in principle, $r_s$ may be distinct for each pair of groups $ij$, yet we omit these indices for clarity. Because molecular simulations deal with interactions of a finite range, we also introduce a "cutting distance" $r_c$; in principle, it can also be different for different pairs of groups. Of course, $r_c > r_s$. With these definitions, the potential pairwise interaction function of RelRes can be compactly presented as follows:

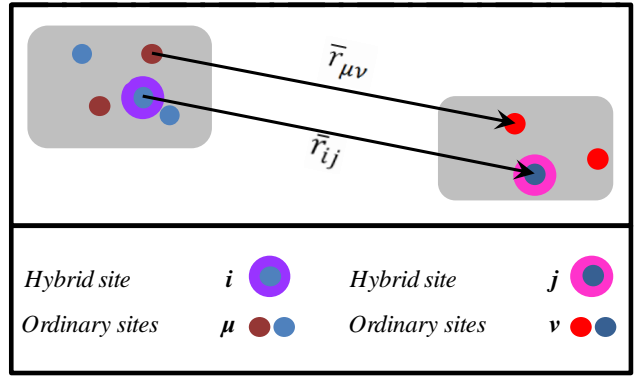

Figure 1. The topology of two groups of atoms. The gray shading just helps us delineate an effective boundary of a group. At a basic level, the sites are represented by disks; the hybrid sites have an extra ring around them, while the ordinary sites lack this aspect. Indeed, the disks signify the FG potentials, while the rings signify the CG potentials. The various colors of these sites signify that they can all have different interaction parameters. The various arrows are distance vectors.

$$\tilde{u}_{\mu\nu}(r) = \begin{cases} u_{\mu\nu}^{FG}(r) - u_{\mu\nu}^{FG}(r_s) + u_{\mu\nu}^{CG}(r_s) - u_{\mu\nu}^{CG}(r_c) & \text{if } r < r_s, \\ u_{\mu\nu}^{CG}(r) - u_{\mu\nu}^{CG}(r_c) & \text{if } r_s \leq r < r_c, \\ 0 & \text{if } r \geq r_c. \end{cases} \quad (1)$$

While $u_{\mu\nu}^{FG}(r)$ formally represents the FG potential for all possibilities of $\mu$ and $\nu$, that is not the case for $u_{\mu\nu}^{CG}(r)$: It is indeed the CG potential for hybrid sites, but it is zero for ordinary sites:

$$u_{\mu\nu}^{CG}(r) = \begin{cases} u_{ij}^{CG}(r) & \mu \wedge \nu \text{ are hybrid} \\ 0 & \mu \vee \nu \text{ is ordinary} \end{cases} \quad (2)$$

The shifting constants at the switching distances, $u_{\mu\nu}^{FG}(r_s)$ and $u_{\mu\nu}^{CG}(r_s)$, as well as the shifting constant at the cutting distance, $u_{\mu\nu}^{CG}(r_c)$, merely provide continuity of the RelRes potential throughout its domain. Note that the complete definition for $\tilde{u}_{\mu\nu}$ is given by eq 10 of ref 33; there, we notably have two smoothing zones, one around $r_s$, as well as one around $r_c$, for purposes of avoiding singularities in the respective force; these smoothing zones are omitted in eq 1 for clarity. Often in our work, we also omit the indices on $\tilde{u}$ for clarity.

While parameterized potentials are often available for FG models, this is not the case for CG models. In order to ensure a correct representation of the system behavior, RelRes proceeds by equating the FG and CG energies at the "infinite limit" of the relative separation between an isolated molecular pair (i.e., $r \gg |\bar{r}_{ij} - \bar{r}_{\mu\nu}|$); this yields the following analytical parameterization[31]

$$u_{ij}^{CG}(r) = \sum_{\mu\nu} u_{\mu\nu}^{FG}(r). \tag{3}$$

Note that the formal derivation of this parameterization proceeds via a multipole series: Equation 3 is just the zero-order term, yet if desired, the first-order and second-order terms are also available for RelRes[32]. Either way, this parameterization is indeed one of the most useful aspects of RelRes, since it is quite rare that the FG and CG potentials can be related analytically.

As in the usual case with molecular simulations, the total energy is given by a pairwise summation over all sites,

$$\tilde{U} = \frac{1}{2}\sum_{i \neq j}\left[\sum_{\mu\nu} \tilde{u}_{\mu\nu}(r_{\mu\nu})\right]. \tag{4}$$

Notice that with $r_s \to \infty$, we obtain a pure FG system, and with $r_s \to 0$, we obtain a pure CG system. We will call the former the "reference" system, as this is frequently the benchmark used in molecular simulations (i.e., it supposedly describes the correct behavior of a liquid of interest). In turn, notice that once the parameterization of eq 3 is employed in eq 4, the total energy of the reference system is approximately captured if $r_s \gg |\bar{r}_{ij} - \bar{r}_{\mu\nu}|$ for all sites (i.e., the dimension of each group is relatively small, in comparison with the switching distance). At this point, it is useful to introduce the discrepancy in the total energy, from the RelRes system to the reference system:

$$\Delta\tilde{U}_{r_s} = \tilde{U}_{r_s} - \tilde{U}_{\infty}, \tag{5}$$

Here, we indicate the dependency on the "switching distance" by the respective index (we will do so throughout if such a distinction is important). Clearly, RelRes is adequate if $\Delta\tilde{U} \to 0$.

## 2.2 RelRes Implementation of the LJ Potential

The RelRes expression of eq 1 can be employed with any FG or CG potential, and if the zero-order term of the multipole series does not vanish, eq 3 is the relevant formula for parameterizing between the FG and CG potentials. Suppose now that all sites intrinsically interact by the LJ potential. In such a case, the FG potential can be fundamentally cast as the following[31]:

$$u_{\mu\nu}^{FG}(r) = 4\epsilon_{\mu\nu}^{FG}\left[\left(\frac{\sigma_{\mu\nu}^{FG}}{r}\right)^{12} - \left(\frac{\sigma_{\mu\nu}^{FG}}{r}\right)^{6}\right] \tag{6}$$

with $\sigma_{\mu\nu}^{FG}$ and $\epsilon_{\mu\nu}^{FG}$ being the respective length and energy parameters of the FG $\mu\nu$ interaction. If we plug this function in eq 3, the CG potential can be sequentially cast as the following[31]:

$$u_{ij}^{CG}(r) = 4\epsilon_{ij}^{CG}\left[\left(\frac{\sigma_{ij}^{CG}}{r}\right)^{12} - \left(\frac{\sigma_{ij}^{CG}}{r}\right)^{6}\right] \tag{7}$$

with $\sigma_{ij}^{CG}$ and $\epsilon_{ij}^{CG}$ being the respective length and energy parameters of the CG $ij$ interaction. Note that because this potential is applicable only beyond $r_s$, the attractive term is dominant in this region, so the repulsive term is almost negligible. The parameterization of eq 3 indeed yields formal relationships between these parameters[31]. We restrict our current work for geometric mixing: If $\epsilon_{\mu\nu}^{FG} = \sqrt{\epsilon_{\mu}^{FG}\epsilon_{\nu}^{FG}}$ and $\sigma_{\mu\nu}^{FG} = \sqrt{\sigma_{\mu}^{FG}\sigma_{\nu}^{FG}}$ for the FG model, then it can be proven that $\epsilon_{ij}^{CG} = \sqrt{\epsilon_{i}^{CG}\epsilon_{j}^{CG}}$ and $\sigma_{ij}^{CG} = \sqrt{\sigma_{i}^{CG}\sigma_{j}^{CG}}$ for the CG model[32]. In turn, the following analytical parameterization ensues between the LJ parameters:

$$\sigma_{l}^{CG} = \frac{\left(\sum_{\nu}\sqrt{\epsilon_{\nu}^{FG}\sigma_{\nu}^{FG\,12}}\right)^{1/3}}{\left(\sum_{\nu}\sqrt{\epsilon_{\nu}^{FG}\sigma_{\nu}^{FG\,6}}\right)^{1/3}}, \quad \epsilon_{l}^{CG} = \frac{\left(\sum_{\nu}\sqrt{\epsilon_{\nu}^{FG}\sigma_{\nu}^{FG\,6}}\right)^{4}}{\left(\sum_{\nu}\sqrt{\epsilon_{\nu}^{FG}\sigma_{\nu}^{FG\,12}}\right)^{2}}. \tag{8}$$

Equation 13 of ref 33 shows the relevant expressions for situation when geometric mixing does not hold. Note, that eq 2 is satisfied by setting the following:

$$\sigma_{\nu}^{CG} = \sigma_{l}^{CG}, \quad \epsilon_{\nu}^{CG} = \begin{cases} \epsilon_{l}^{CG} & \nu \text{ is hybrid} \\ 0 & \nu \text{ is ordinary} \end{cases} \tag{9}$$

This RelRes algorithm, as presented by eqs 6-9, is currently implemented in LAMMPS: All details can be found in ref 33 in the context of eq 20, as well as in the LAMMPS documentation itself[39]; in the current work, we just present a brief description. Using the terminology of LAMMPS, this variant of RelRes can be specifically used as `lj/relres` in the `pair_style` command[33]. Notably, `lj/relres` is programmed as an extension of `lj/smooth`. Fundamentally, `lj/smooth` deals with one potential: In its `pair_style` command, it takes two parameters for the cutting distance (where it starts smoothening, and where it finishes smoothening), and in its `pair_coeff` command, it takes two parameters representing the LJ energy and length scales of each site type. Conversely, `lj/relres` deals with two potentials: In its `pair_style` command, it takes four parameters in total for $r_s$ and $r_c$ (where each starts smoothening, and where each finishes smoothening), and in its `pair_coeff` command, it takes four parameters in total representing $\{\epsilon^{FG}, \sigma^{FG}, \epsilon^{CG}, \sigma^{CG}\}$ of each site type. Note that in this LAMMPS implementation, the smoothing zones, which guarantee the continuity of the force around $r_s$ and around $r_c$, are characterized by polynomial functions.

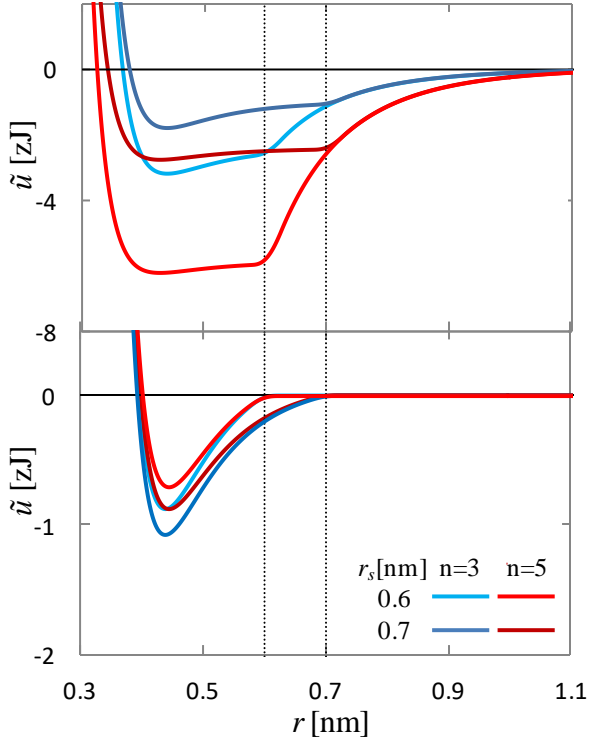

Figure 2. The RelRes potential function. While the potential between hybrid sites is shown in the top panel, the potential between ordinary sites is shown in the bottom panel. There are actually four separate examples in each panel: The blue curves represent propane (mapping ratio of 3), and the red curves represent neopentane (mapping ratio of 5); the light colors are for the switching distance of 0.6nm, while the dark colors are for the switching distance of 0.7nm. Each dotted vertical line corresponds with the location of the switching distance.

Figure 2 shows four variations of this RelRes potential in LAMMPS. Here, propane, with its mapping ratio of 3, is plotted in blue, and neopentane, with its mapping ratio of 5, is plotted in red. The darker colors are used for the longer switching distance ($r_s = 0.7$nm), and the lighter colors are used for the shorter switching distance ($r_s = 0.6$nm). The potential between a pair of hybrid sites is plotted in the top panel, while the potential between a pair of ordinary sites is plotted in the bottom panel; notice that the ordinate of the former is higher by a fourfold factor.

For the top panel, it can be clearly observed how the FG potential is replaced by the CG potential at $r_s$, with the continuity of the hybrid potential at that location being ensured by the shifting constants $u^{CG}(r_s)$ and $u^{FG}(r_s)$. This is of course not the situation for the bottom panel, as by construction, the ordinary potential vanishes beyond $r_s$. Besides, in each panel, there are obvious discrepancies between the four variations that we show. For better understanding these discrepancies, suppose for now that the parameters across all sites are identical in their values, which in turn yields for eq 8 $\sigma_l^{CG} = \sigma_v^{FG}$, together with $\epsilon_l^{CG} = n^2 \epsilon_v^{FG}$. In reality, the parameters of the sites for propane and for neopentane (especially the energy ones) are somewhat different: Plugging in actual values, we find that $\sigma_l^{CG} \approx \sigma_v^{FG}$ for both alkanes, together with $\epsilon_l^{CG} \approx 8\epsilon_v^{FG}$ for propane and $\epsilon_l^{CG} \approx 21\epsilon_v^{FG}$ for neopentane. Thus in the top panel, the main reason that the LJ well deepens with an increasing mapping ratio (from blue curves to red curves) is the perfector of $\sim n^2$, which eq 8 hides; concurrently, the main reason that the LJ well deepens with a decreasing switching distance (from dark curves to light curves) stems in the shifting constants $u^{CG}(r_s)$ and $u^{FG}(r_s)$. At the same time in the bottom panel, the curves are almost the same since the parameters across all propane and neopentane ordinary sites are quite similar. However, since $\sigma_v^{FG}$ and $\epsilon_v^{FG}$ are not absolutely the same, there are still slight differences between the red and blue curves (obviously, the mapping ratio has no role in the bottom panel). At the same time, just as we mentioned in the context of the top panel, the slight differences between the light and dark curves stem in the shifting constant.

## 2.3 Kullback–Leibler Entropy

We observed that RelRes, with the appropriate choice of $r_s$, can successfully retrieve the various static and dynamic features of several alkane systems[33]. Still, given an arbitrary liquid, it becomes of interest in analyzing the corresponding multiscale errors in a systematic and robust manner. The task of examining the myriad of errors associated with structural correlations, thermal properties, etc. is indeed daunting. Luckily, it has been exhaustively shown that the informatic measure of the KL Entropy can collectively signal many errors associated with multiscale simulations[37].

On a fundamental level in information theory, the KL Entropy enumerates the amount by which a "mimic" probability distribution deviates off a "reference" probability distribution (it is indeed an asymmetric measure). In molecular simulations of multiscale systems, the KL Entropy was formulated as a functional of Boltzmann distributions, originally applied from a pure FG model to a pure CG model[35]. In the context of RelRes, we slightly modify this expression, so that the KL entropy measures how much a RelRes system (which formally uses $r_s$) deviates off its reference system (which effectively has $r_s \to \infty$):

$$S_{KL,r_s} = \int p_\infty(\Gamma) \ln\left[\frac{p_\infty(\Gamma)}{p_{r_s}(\Gamma)}\right] d\Gamma, \qquad (10)$$

where $p$ denotes the ensemble probability of a multidimensional configuration $\Gamma$, and the integration is done over the entire space of the reference system. Realize that in the context of RelRes, no mapping operator is required in this expression since all degrees of freedom are maintained. Throughout most of our work, we omit the index $r_s$ of $S_{KL}$.

Information theory offers other metrics between probability distributions[40]. For example, it is often suggested that the symmetric measure of the Jensen-Shannon Entropy is the most fundamental one: It is the average of two KL entropies, with the roles of the mimic and reference systems switched between the two KL entropies[41]. This metric indeed has some advantages over the KL Entropy (e.g., it eliminates the possibility of a singularity in the case for which $p_{r_s}(\Gamma) = 0$). However, the KL Entropy has the main advantage in the simplicity of its calculation. It solely requires conducting a molecular simulation of the reference system, and there is no need for ever creating any molecular simulations with a finite $r_s$. If one is interested in the Jensen-Shannon Entropy for many $r_s$, one must perform a separate molecular simulation for each $r_s$.

If we plug in the canonical probabilities in eq 10, we obtain the following expression[35]:

$$S_{KL} = \beta \langle \tilde{U}_{r_s} - \tilde{U}_\infty \rangle_\infty - \beta[A_{r_s} - A_\infty], \qquad (11)$$

where $\beta$ is the inverse temperature, $A$ is the free energy of a certain system, and the averaging is performed in the canonical ensemble of the reference system (indicated by infinity). Again, realize that unlike in ref 35, there is no mapping entropy here, since RelRes maintains all degrees of freedom. Note that since free energies are computationally cumbersome, this expression is not very convenient for calculation via molecular simulations.

Using Zwanzig perturbation[42], eq 11 can be reformulated as follows[37]:

$$S_{KL} = \ln\langle\exp[\beta(\Delta\widetilde{U} - \langle\Delta\widetilde{U}\rangle_\infty)]\rangle_\infty, \quad (12)$$

where $\Delta\widetilde{U}$ is of course dependent on the switching distance, as defined by eq 5. This expression presents the KL Entropy as a special measure of the fluctuations in the Hamiltonian discrepancy $\Delta\widetilde{U}$. This formula is quite convenient on a computational level since it just involves averages of configurational functions (there are no free energies here)[38]. In such a manner, the KL Entropy can be evaluated at once by performing a molecular simulation of the reference system. In fact, just with a single run of such a molecular simulation, we can readily obtain the values of many $S_{KL,r_s}$ (each having its own switching distance).

Since RelRes typically captures the reference system very well, we expect that $\Delta\widetilde{U}$ approaches zero, and in turn, we can perform a Taylor approximation for eq 12[37]:

$$\check{S}_{KL} = \frac{1}{2}\beta^2(\langle\Delta\widetilde{U}^2\rangle_\infty - \langle\Delta\widetilde{U}\rangle_\infty^2). \quad (13)$$

This is naturally based on an approximation, as compared with eq 12, and for emphasizing this aspect, we have the "breve" above this KL Entropy. This formula suggests that the KL Entropy is roughly proportional with the variance of the Hamiltonian discrepancy $\Delta\widetilde{U}$. Importantly, this expression has a tremendous advantage in that it involves no exponentials in its averaging (these yield significant statistical errors)[38]. Moreover, we can readily obtain the values of many $\check{S}_{KL,r_s}$ (each having its own switching distance) with a single run of such a molecular simulation of the reference system.

These two formulae give two estimates for the KL Entropy (ideally, both yield the same value). However, the convergence of eq 12 can be problematic as extreme values of $\Delta\widetilde{U}$ can dominate the average over the exponential function. Even though $\Delta\widetilde{U}$ approaches zero in RelRes (especially with a sufficient choice for $r_s$), a considerable amount of configurations must be sampled, so that the statistical error on the KL Entropy is moderate[38]. Note that eq 13 does not have such drawbacks because of the absence of an exponential function[38].

Moreover, the KL Entropy is an extensive property that naturally depends on the size of the system. Thus, we define an associated intensive property, the KL Entropy per group of sites: $s_{KL} = S_{KL}/\mathcal{N}$ and $\check{s}_{KL} = \check{S}_{KL}/\mathcal{N}$, where $\mathcal{N}$ is the number of groups in the system (the number of groups can be different than the number of molecules $N$). As an emphasis, the extensive KL Entropies are given by the upper case letters, yet the intensive KL Entropies are given by the lower case letters. In our work, we will present just these intensive metrics, and we will colloquially call $s_{KL}$ the exact value of the KL Entropy (always calculated utilizing eq 12) and $\check{s}_{KL}$ the approximate value of the KL Entropy (always calculated utilizing eq 13); data for the extensive metrics is never shown.

## 2.4 Setup of the Molecular Simulations

The setup of our molecular simulations is almost identical to the one in ref 33. Here, we succinctly mention the most important distinctions. Depicted in Figure 3, there are four different alkanes that we study:

- Two monomers, $C_3H_8$ (Propane) and $C_5H_{12}$ (Neopentane). Each has one group per molecule, $n = 3$ for propane and $n = 5$ for neopentane.
- Two dimers, $C_3H_7$-$C_3H_7$ (Hexane) and $C_5H_{11}$-$C_5H_{11}$ (Bineopentyl). Each has two groups per molecule, $n = 3$ for hexane and $n = 5$ for bineopentyl.

We indeed perform all molecular simulation in LAMMPS: The reference systems employ `lj/smooth` pair style, yet the RelRes system employ `lj/relres` pair style. Regarding the intermolecular energetics, the respective FG models utilize the OPLS_UA parameters[43], and the respective CG models are correspondingly parameterized by eq 8 (geometric mixing is applicable throughout).

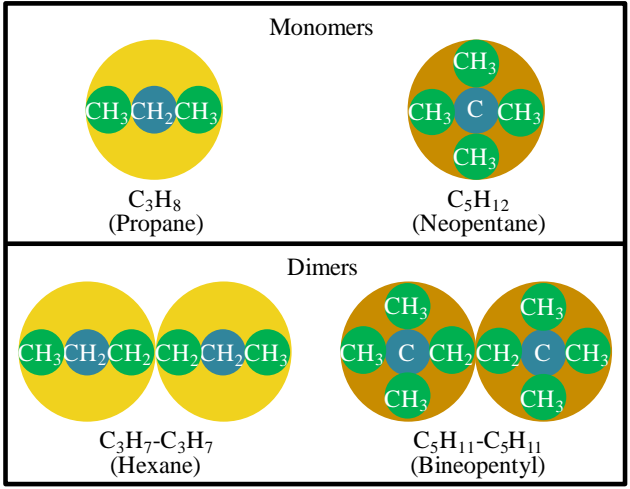

Figure 3. The alkane molecules with their mapping. The orange circles delineate an effective boundary of a group; a light orange is for $n = 3$ (on the left), and a dark orange is for $n = 5$ (on the right). Actual atoms are given in green: "Hybrid" sites are represented by dark green circles, while "ordinary" sites are represented by light green circles. The top panel focuses on the two monomers $C_3H_8$ and $C_5H_{12}$. The bottom panel presents the two dimers $C_3H_7$-$C_3H_7$ and $C_5H_{11}$-$C_5H_{11}$.

All systems use a cutting distance of 1.3nm, while the switching distance in the RelRes systems is varied from 0.5nm to 0.8nm (at steps of 0.1nm); the smoothing zones that we use in this work are analogous with those in ref 33. Regarding the intramolecular energetics, all systems are based on the AMBER_UA parameters[44].

All systems which we examine are uniform (single-component and single-phase) fluids. All data, which we report in this work, is collected by molecular simulations in the canonical ensemble: Besides having a constant $N$, they also naturally have a fixed temperature and a fixed volume. Still, the molecular simulations also involve a preliminary step in the isothermal-isobaric ensemble (having a fixed temperature and a fixed pressure): The sole purpose of this preliminary step is for setting, for the specific pressure, the proper volume of each alkane system.

In all cases, a box with periodic boundaries is utilized. In general, we examine four different system sizes ($\mathcal{N} = \{2000, 3456, 5488, 16000\}$), yet we most often focus on the smallest and largest sizes ($\mathcal{N} = \{2000, 16000\}$). Most molecular simulations have a temperature of $T = 290\text{K}$ (equivalent with a thermal energy of 4.0zJ) and a pressure of $P = 15000$ Torr (equivalent with an energetic density of $2.0\text{MJ}/\text{m}^3$). We call these the "standard conditions". In addition, we investigate the behavior across state space via two separate sets of molecular simulations. In the first set, we run four temperatures that are higher than the standard one ($T = \{435\text{K}, 580\text{K}, 870\text{K}, 1160\text{K}\}$); all these molecular simulations maintain the standard density of the alkane system in focus. In the second set, we run four densities that are lower than the standard one (these are correspondingly scaled down by factors of $\{4.0, 8.0, 16.0, 32.0\}$, and it is practically done by increasing the volume of the box accordingly for each alkane system as compared with its standard volume); all these molecular simulations are performed at the nonstandard temperature of 1160K (the highest one).

Each reference system is prepared at standard conditions via a preliminary step in the isothermal-isobaric ensemble for 6.0ns with a timestep 4.0fs; this fixes the standard volume (also standard density) for each alkane system, and it also notably becomes the basis for all RelRes simulations. Then, for each system, an equilibration is done in the canonical ensemble: For standard conditions, it is performed for 1.0ns with a timestep $4.0\text{fs}$, yet for nonstandard conditions, it is performed for 2.0ns with a timestep 4.0fs. After the equilibration, data is ultimately collected via molecular simulations also in the canonical ensemble for 1.0ns with a timestep 1.0fs. We emphasize that all of our systems are single-component and single-phase fluids of nonpolar alkanes, and this allows us for achieving sufficient statistics with relatively short runtimes of fairly large timesteps.

For calculating the KL Entropy, the trajectories of just the reference systems are processed (of all four sizes). Every 200 timesteps, $\Delta \widetilde{U}_{r_s}$ is evaluated for each RelRes potential (with its specific switching distance). Then, by running averages over all the trajectories, with a total of 1000 samples, the exact $s_{KL}$ is computed via eq 12, and the approximate $\check{s}_{KL}$ is computed via eq 13; note that in a preliminary exploration, we actually vary the number of samples used in the computation of the KL Entropy, and, in turn, we prove for ourselves that 1000 samples is totally sufficient for achieving convergence. Besides for enhancing the statistics, we execute ten replicates of these molecular simulations. We accordingly calculate the standard errors of the exact and approximate KL Entropies based on a 0.95 confidence interval.

For analyzing the multiscale errors associated with thermal properties, structural correlations, etc., the trajectories of each RelRes system, of each switching distance ($r_s = \{0.5, 0.6, 0.7, 0.8\}$nm), are also processed (just of the smallest and largest $\mathcal{N}$). For the comparison of the structural behavior, we focus on the radial distribution $g(r)$ between hybrid sites (it was shown in ref 33 that RelRes mostly impacts this $g(r)$). This radial distribution is sampled every 10 timesteps with averaging over 50000 samples. We specifically cast a functional for comparing the RelRes $g(r)$ with the reference $g(r)$:

$$s_{KL} = \int_0^{r_c} p_\infty(r) \ln\left[\frac{p_\infty(r)}{p_{r_s}(r)}\right] r^2 dr. \tag{14}$$

Here, $p(r) = g(r)/\int g(s)s^2 ds$ is the probability density associated with the radial distribution. In analogy with eq 10, one may think of this functional as a unidimensional version of the KL Entropy, which is specific for radial distributions. For comparison of the thermal behavior, we use, in analogy with ref 33, the (normalized) multiscale error in the intermolecular potential energy, as well as the (normalized) multiscale error in the virial associated with it. Here is how we define these multiscale errors for their respective arithmetic means and standard deviations:

$$\left[\langle\widetilde{U}\rangle\right]^*_{\text{err}} = \frac{\langle\widetilde{U}_{r_s}\rangle - \langle\widetilde{U}_\infty\rangle}{\langle\widetilde{U}_\infty\rangle}, \quad \left[\langle\delta\widetilde{U}^2\rangle^{1/2}\right]^*_{\text{err}} = \frac{\langle\delta\widetilde{U}^2_{r_s}\rangle^{1/2} - \langle\delta\widetilde{U}^2_\infty\rangle^{1/2}}{\langle\delta\widetilde{U}^2_\infty\rangle^{1/2}}, \tag{15}$$

$$\left[\langle\widetilde{W}\rangle\right]^*_{\text{err}} = \frac{\langle\widetilde{W}_{r_s}\rangle - \langle\widetilde{W}_\infty\rangle}{\langle\widetilde{W}_\infty\rangle}, \quad \left[\langle\delta\widetilde{W}^2\rangle^{1/2}\right]^*_{\text{err}} = \frac{\langle\delta\widetilde{W}^2_{r_s}\rangle^{1/2} - \langle\delta\widetilde{W}^2_\infty\rangle^{1/2}}{\langle\delta\widetilde{W}^2_\infty\rangle^{1/2}}, \tag{16}$$

The data for calculating arithmetic means and standard deviations is collected every 200 timesteps over the entire molecular simulation (there are 5000 samples in total).

## 3 Results

As already mentioned, our ultimate goal is to determine if the KL Entropy can be used as a general metric for measuring the adequacy of our hybrid algorithm: Namely, does the value of the KL Entropy correspond well with most multiscale errors (in structural correlations, thermal properties, etc.) between the RelRes system and the reference system? We specifically do such an examination for the alkane systems mentioned above. Here, we also analyze the dependency of the KL Entropy on the system size, and we do so for both the exact and approximate values of the KL Entropy. Finally, we also look at the behavior of the KL Entropy as a function of state space.

### 3.1 Variation of System Size

As already mentioned, achieving convergence for eq 12 (yet not for eq 13) is problematic because of the exponential function in its average. Thus, before varying the system size, we must ensure that we can converge on the correct values of the KL Entropy with a finite number of configurations. As such, we examined the dependency of the KL entropy on the number of samples (analogously with the analysis performed in ref 38, as shown in its Figure 5). In this provisional study, we just examined propane and neopentane, with their recommended $r_s$[33]; besides, we employed the largest system ($\mathcal{N} = 16000$) for eq 12 and the smallest system ($\mathcal{N} = 2000$) for eq 13 (these systems are most representative of our results in this current work). Consequently, we noticed that using 200-500 samples in both the exact and approximate calculations is typically sufficient for converging on the appropriate values of the KL Entropies (for brevity, these plots are not shown here). We thus confirm that sampling 1000 configurations of our alkane systems in calculating the KL Entropy is wholly sufficient for achieving convergence.

We now begin investigating the dependency of the system size on the exact and approximate versions of the KL Entropy, for all our four alkanes. Figure 4 summarizes the results of such an analysis, with the inverse size of the system plotted on the abscissa, while both variants of the KL entropy are plotted on the ordinate. We focus on just a few specific scenarios in the four panels: Each panel is for a different alkane, as well as for a different $r_s$. The exact KL Entropy is plotted in red, and the approximate KL Entropy is plotted in blue. The solid filled markers represent actual data, which are obtained via the averaging in eqs 12 or 13; conversely, the empty hollow markers are obtained by extrapolation to systems of infinite size ($\mathcal{N} \to \infty$), using linear regression that is fitted for all data of finite size.

Figure 4 notably shows the standard error associated with the calculation of each value of the KL Entropy. As mentioned earlier, each error bar corresponds with a 0.95 confidence interval. Notice that the error bars do not depend much on the system size; therefore, the standard errors for the extrapolated infinity size are calculated just by averaging the standard errors of all the finite sizes. Also, observe that in general, error bars are larger for higher values of the KL Entropy: Reference 38 indeed exemplified that the standard errors for both the exact and approximate calculations of the KL Entropy (eqs 12 and 13, respectively) are almost proportional with the values of the KL Entropy themselves. Besides in most cases, the error bars are almost the same size as the markers themselves (realize that in Figure 4, the markers are made especially smaller); as such, in all subsequent plots of these values of the KL

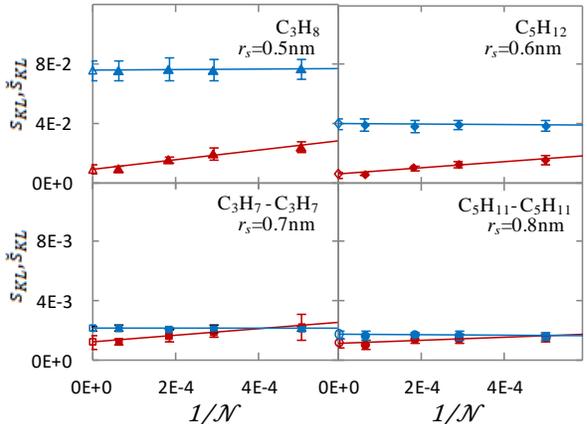

Figure 4. The KL Entropy as a function of the inverse size of the system. The exact value is shown in red, and the approximate value is shown in blue. Filled markers correspond with actual data attained by averaging, as given by eqs 12 or 13; conversely, hollow markers (at the zero value of the abscissa) are achieved by extrapolation for infinite size using linear regression. Each panel is for a different alkane: The top panels show the monomers, and the bottom panels show the dimers; the left panels are for the mapping ratio of 3, and the right panels are for the mapping ratio of 5. Each panel employs its own switching distance. Notice the tenfold difference in scale between the top and bottom panels. Here, the markers are made especially small so that the error bars are relatively noticeable.

Entropy, the markers are made somewhat larger, and we completely omit presenting any of these error bars.

In ref 33, we found that the recommended switching distances are $r_s^* = 0.6$nm for $n = 3$ and $r_s^* = 0.7$nm for $n = 5$. The top panels intentionally have $r_s$ slightly lower than recommended, and the bottom panels intentionally have $r_s$ slightly higher than recommended. Increasing the switching distance naturally improves the performance of RelRes, which in turn decreases the value for the KL Entropy. Thus, Figure 4 gives us a comprehensive view of the KL Entropy, with noticeably higher values in the top panels and noticeably lower values in the bottom panels (they roughly differ by an order of magnitude).

The overall results are conceptually similar for all our alkanes. In fact, the most important aspect of Figure 4 is the striking distinction in the behavior between the approximate and exact variants of the KL Entropy. The approximate KL Entropy almost does not depend on the size of the system (it is essentially constant in the graph), but the exact KL Entropy is inversely proportional with the system size (it is essentially linear in the graph). Another notable observation is that $š_{KL}$ gives an excellent estimate for $s_{KL}$ for the systems with the low values in the bottom panels, which is clearly not the case for the high values in the top panels.

For better understanding the correspondence between the exact and approximate KL Entropies, we compare $s_{KL}$ and $š_{KL}$ for each specific $\mathcal{N}$ in Figure 5. In particular, we plot the exact KL Entropy on the abscissa and the approximate KL Entropy on the ordinate, with each system size having its unique color (notice that besides the finite sizes, we also show here the infinite limit). Realize that all of our alkanes are in Figure 5, with each one having a unique shape for its marker; notably for each alkane, we present all four switching distances (this is unlike in Figure 4).

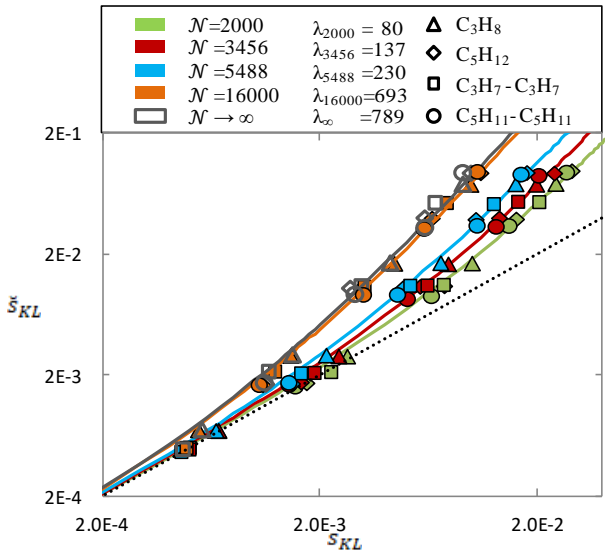

Figure 5. The relationship between the exact (abscissa) and approximate (ordinate) values of the KL Entropy. Both axes are in a logarithmic scale. The different system sizes are presented with different colors, while the different alkanes are presented with different marker shapes. Solid lines indicate a quadratic regression, as given by eq 17, fitted by a tuning factor $\lambda_\mathcal{N}$. The dotted black line is the identity function. Just like in Figure 4, the filled markers are for actual data of finite systems, while hollow markers represent extrapolation for infinite systems. No error bars are shown here as they are the same as those given in Figure 4; since the markers here are made rather large, they essentially encapsulate their corresponding error bars.

Figure 5 suggests that we can describe the approximate KL Entropy as a function of the exact KL Entropy: $\check{s}_{KL} = f(s_{KL})$. In consideration of the derivation from eq 12 to eq 13, the ideal limit has $\check{s}_{KL} \cong s_{KL}$, and this becomes the first-order Taylor approximation for our function of interest: It is specifically given by the dotted black line, which obviously passes through the origin with a unity slope. This linear function is clearly valid for $s_{KL} \leq 5 \cdot 10^{-4}$, as observed in the bottom left corner of Figure 5. For $s_{KL} \geq 5 \cdot 10^{-4}$, further terms in the Taylor expansion obviously have a significant influence. Here, we just employ the second-order Taylor approximation:

$$\check{s}_{KL} \cong s_{KL} + \lambda_{\mathcal{N}} s_{KL}^2 \tag{17}$$

with $\lambda_{\mathcal{N}}$ being a fitting parameter for each specific system size $\mathcal{N}$: We fit this quadratic function for each data set by correspondingly minimizing the sum of the squared errors of their logarithms while varying $\lambda_{\mathcal{N}}$; the coefficient of determination ranges from 0.99 to 1.00 across the various system sizes. In turn, we plot solid colored lines for all system sizes in Figure 5. We thus deem that no further polynomial terms are required in the Taylor expansion for describing our function of interest: Equation 17 is much sufficient.

Summarizing our current observations, while complementing some of our earlier statements, the approximate KL Entropy indeed proves as the more convenient metric for use than the exact KL Entropy. We already mentioned this aspect in the context of eqs 12 and 13: The exact expression involves averaging over exponentials, which is prone for significant statistical uncertainty. Now, we also observe that the approximate expression is independent of the size of the system. With the use of eq 17, we can now devise a convenient way for estimating the value of the exact KL Entropy, by just having knowledge of the value of the approximate KL Entropy. Thus, we invert eq 17:

$$s_{KL} \cong \frac{\left(\sqrt{4\lambda_{\mathcal{N}} \check{s}_{KL} + 1} - 1\right)}{2\lambda_{\mathcal{N}}}. \tag{18}$$

In principle, this expression applies for the same $\mathcal{N}$ of both the exact and approximate KL Entropies. However, we noted in Figure 4 that the approximate value is almost independent of system size, so even for the smallest system, $š_{KL}(\mathcal{N} = 2000) \cong š_{KL}(\mathcal{N} \to \infty)$. Besides, we can also notice in Figure 4 that the exact value of the largest system almost reaches the infinite limit, $s_{KL}(\mathcal{N} = 16000) \cong s_{KL}(\mathcal{N} \to \infty)$. So, for the remainder of our work, we will just continue examining two system sizes: Most importantly, we will calculate the approximate KL Entropy for the smallest systems ($\mathcal{N} = 2000$), and we will calculate the exact KL Entropy for the largest systems ($\mathcal{N} = 16000$). In turn, we define an ideal parameter $\lambda^*$, which corresponds with the best fit for eq 18 that links $s_{KL}$ of the "infinite" system ($\mathcal{N} = 16000$) with $š_{KL}$ of the "infinitesimal" system ($\mathcal{N} = 2000$).

## 3.2 Multiscale Errors in Structural Correlations

We now compare the structural correlations of the RelRes systems with those of the reference systems, and we do so with the aid of the KL Entropy. In our current work, we just focus on the radial distributions between pairs of hybrid sites. This is because in our previous study,[33] we found that RelRes has the most adverse impact on those $g(r)$; all $g(r)$ that involve ordinary sites remain essentially unaltered with those of the reference system.[33]

As such, we start by presenting these essential radial distributions in Figure 6, with each panel corresponding with one of the four alkanes. Each panel presents the reference system as the black solid line, together with three variations for the RelRes system: The one, which uses the recommended switching distance, is given by the purple dashed line, while the other two, which use larger and smaller switching distances, are given by the red and blue dotted lines, respectively. Our observations here are complementary with those of our previous study[33]: Increasing $r_s$ improves the capability of RelRes in capturing the structural correlations.

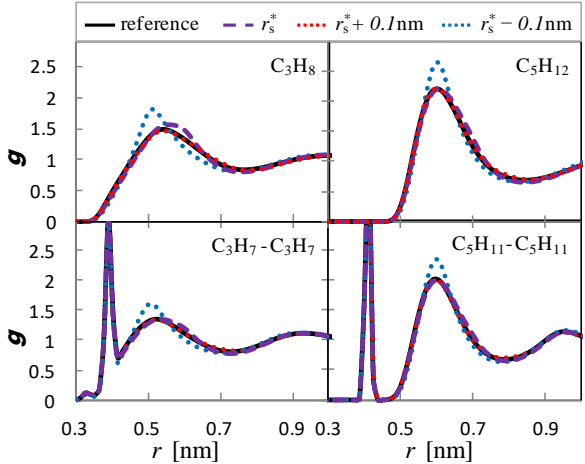

Figure 6. The radial distribution, between hybrid sites, as a function of their pairwise distance. Consistent with Figure 4, each panel is for a specific alkane: The top panels present monomers, and the bottom panels present dimers; the left panels are for a mapping ratio of 3, and the right panels are for a mapping ratio of 5. In each panel, the reference system is shown as a black solid line, while the RelRes systems are given by colored lines: The purple dashed line is for the recommended switching distance ($r_s = r_s^*$), while the red and blue dotted lines are respectively for larger and smaller switching distances ($r_s = r_s^* \pm 0.1\text{nm}$). Recall that according to ref 33, $r_s^* = 0.6\text{nm}$ for $n = 3$ and $r_s^* = 0.7\text{nm}$ for $n = 5$.

For performing a rigorous analysis of such structural correlations via the KL Entropy, a functional is required for the radial distribution. Indeed, $s_{KL}$ was defined for such purposes in eq 14, and this functional becomes our multiscale error of focus here. Figure 7 plots this structural functional in terms of the KL Entropy for all of our alkanes. The left panel, which is for systems of size 2000, employs the approximate value of the KL Entropy (calculated via eq 13), and the right panel, which is for systems of size 16000, employs the exact value of the KL Entropy (calculated via eq 12). Analogous with our earlier observations, notice that in Figure 7, the exact values are generally smaller than their corresponding approximate values: While the same scale is used for the abscissa, the data in the right panel is squeezed to its left portion.

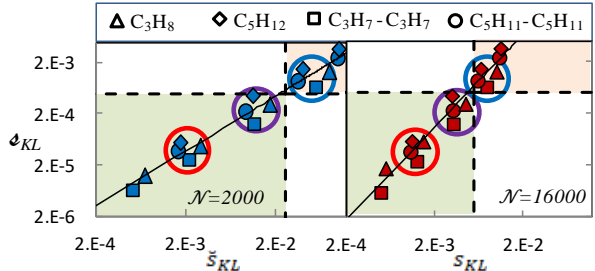

Figure 7. The correspondence of the KL Entropy (abscissa) with the structural functional of eq 14 (ordinate). Both axes are in a logarithmic scale. The left panel presents molecular simulations of size 2000, with the approximate value of the KL Entropy being measured here. The right panel presents molecular simulations of size 16000, with the exact value of the KL Entropy being measured here. Different alkanes are represented by markers of different shapes, as shown in the legend. Black solid lines represent fits of power laws (the coefficient of determination is 0.96 for $\breve{s}_{KL}$ and 0.95 for $s_{KL}$). There are four quadrants in each panel (with their boundaries marked by dashed lines): The lower left (green) quadrant is for "good" data ($s_{KL} < 5 \cdot 10^{-4}$), and the upper right (orange) quadrant is for "bad" data ($s_{KL} > 5 \cdot 10^{-4}$); regarding the KL entropy, the boundaries are given by $\breve{s}_{KL}^* = 0.0250$ and $s_{KL}^* = 0.0055$. Besides, we circle data (in groups of four) for emphasizing the correspondence with Figure 6 (they are color coded in the same manner): Purple circles are for molecular simulations that use $r_s^*$, while red and blue circles are for molecular simulations that use $r_s^* + 0.1$nm and $r_s^* - 0.1$nm, respectively.

The various markers correspond with results of molecular simulations of different alkanes, and we fit a power law in each panel, given by a black solid line ($s_{KL} \propto \breve{s}_{KL}^{1.08}$ and $s_{KL} \propto s_{KL}^{1.79}$). The fit is excellent in both panels, with the coefficient of determination being 0.96 on the left side and 0.95 on the right side. This means that both the exact and approximate KL Entropies are fantastic indicators of the multiscale error associated with structural correlations. As an extra analysis, we divide each panel into four quadrants so that we can categorize the adequacy of replication of the structural behavior: All molecular simulations with a "good" representation are in the bottom left (green) quadrant, and all molecular simulations with a "bad" representation are in the top right (orange) quadrant; note that in Figure 7, no data is located in the other two quadrants, which further reiterates the efficacy of the KL Entropy in signaling structural behavior. Realize that the boundaries of these quadrants were picked in such a way that all four molecular simulations with the recommended switching distances are located just at the periphery of the appropriate quadrant. This corresponds with $s_{KL}^* = 0.0055$ and $\check{s}_{KL}^* = 0.0250$ for the KL Entropies, while regarding the structural behavior, $\mathcal{s}_{KL}^* = 5 \cdot 10^{-4}$. Of course, the choice of our boundaries is not unique: If desired, others can establish lower or higher boundaries for the KL Entropy in order to retrieve lower or higher discrepancies in structural representation.

### 3.3 Multiscale Errors in Thermal Properties

Now, we turn to analyze the relationship of the KL Entropy with the corresponding multiscale errors in thermal properties. In Figure 8, we specifically examine the errors in the intermolecular potential energy as defined by eq 15. Errors in the arithmetic means are presented in the top panels, and errors in the standard deviations are presented in the bottom panels. Similar with Figure 7, the left panels show small systems (size 2000), which employ the approximate values of the KL Entropy, and the right panels show large systems (size 16000), which employ the exact values of the KL Entropy.

There is a decent relationship of the KL Entropy with both errors: In general, lower values of the KL Entropy go together with lower values of both errors. However, the relationship is clearly not as strong as the one in Figure 7 (we thus make no attempt at making a fit here). However, we still partition into quadrants. Analogous with Figure 7, we again use the same vertical boundaries, $\check{s}_{KL}^* = 0.0250$ and $s_{KL}^* = 0.0055$, whilst invoking a horizontal boundary of 5 percent for both the mean and deviation of the potential energy (i.e., $[\ldots]_{err}^* = 0.05$). Importantly, most molecular simulations that use $r_s \geq r_s^*$ (red ovals, as well as purple ones) appear in the lower left (green) quadrant, meaning that they yield "good" replication, but those that use $r_s < r_s^*$ (blue ovals) often appear in the upper right (orange) quadrant, meaning that they yield "bad" replication. Besides, there are exceptions for this rule, as we notice a few samples in the upper left and lower right quadrants (remember that in Figure 7, those quadrants are empty).

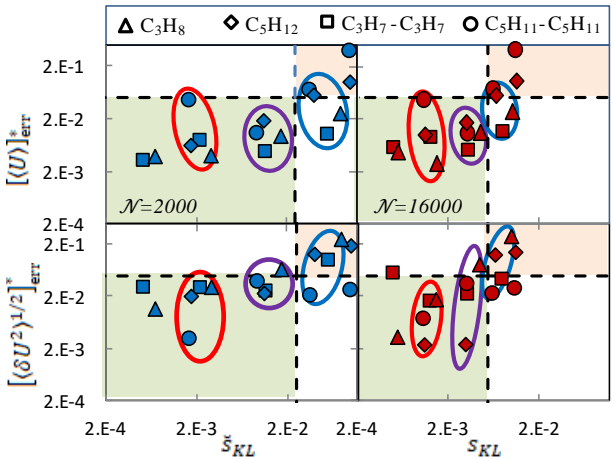

Figure 8. The dependency of the KL Entropy (abscissa) with the normalized multiscale errors of the intermolecular potential energy (ordinate). Both axes are in a logarithmic scale. The top panels show the errors in the means, while the bottom panels show the errors in the deviations. The left panels present data for small systems of 2000 groups (employing the approximate KL Entropy), and the right panels present data for large systems of 16000 groups (employing the exact KL Entropy). Different alkanes are represented by markers of different shapes, as shown in the legend. Partitioning into quadrants (bounded by dashed lines) is done with the vertical boundaries of $\check{s}_{KL}^* = 0.0250$ and $s_{KL}^* = 0.0055$, while the horizontal boundaries are set at $[...]_{err}^* = 0.05$. The purple ovals contain molecular simulations using $r_s^*$, while the red and blue ovals contain molecular simulations using $r_s^* + 0.1$nm and $r_s^* - 0.1$nm, respectively.

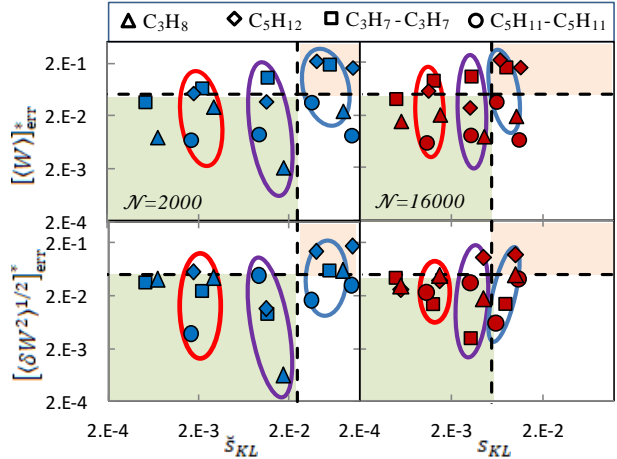

Figure 9. The dependency of the KL Entropy (abscissa) with the normalized errors of the intermolecular virial force (ordinate). All the formatting here is equivalent with Figure 8.

Figure 9 is essentially equivalent with Figure 8, except that it focuses on the intermolecular virial force (rather than on the intermolecular potential energy), with the corresponding multiscale errors defined by eq 16. For all practical purposes, our observations for Figure 9 are more or less analogous with those of Figure 8, yet the relationship here is somewhat weaker.

### 3.4 Investigation across State Space

Everything that we have shown up until now has been conducted at a single temperature of $T = 290K$ and a single pressure of $P = 15000$ Torr: At this point, we are convinced that the KL Entropy (both the approximate and exact variants) is an excellent metric for measuring multiscale errors in structural correlations and a decent metric for measuring multiscale errors in thermal properties. As such, we can now proceed further in state space, examining the behavior of the KL Entropy across multiple temperatures and densities.

Figures 10 and 11 present the ensuing results across state space for all alkanes (specifically those that use the recommended $r_s^*$). Just as we have done previously, $\check{s}_{KL}$ is given for systems with $\mathcal{N} = 2000$, and $s_{KL}$ is given for systems with $\mathcal{N} = 16000$; in all panels, both variants of the KL Entropy are jointly plotted on the ordinate. Regarding the abscissa, Figure 10 shows the dependency on the inverse temperature, and Figure 11 shows the dependency on the group density. In all cases, we respectively fit power laws that are relevant for the limits $\beta \to 0$ and $\varrho \to 0$. Noteworthy, everything in Figure 10 is at the standard density of each alkane, whilst everything in Figure 11 is at the nonstandard temperature of 1160K (this guarantees that all systems are stable, meaning that they are single-component and single-phase fluids).

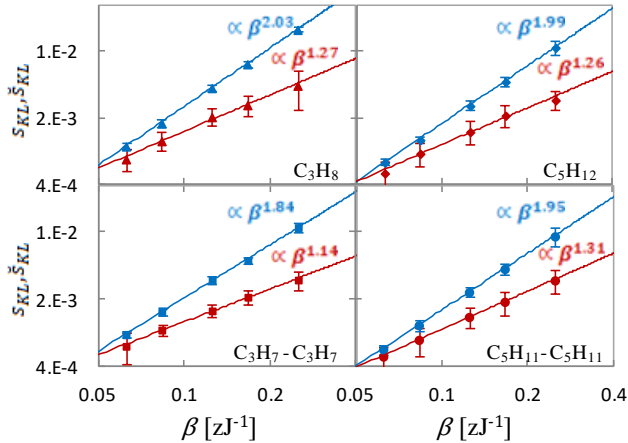

Figure 10. The KL Entropy as a function of the inverse temperature. All axes are scaled logarithmically. Each panel presents a specific alkane as indicated by the corresponding label. The top panels are for the monomers, and the bottom panels are for the dimers; the left panels are for $n = 3$, and the right panels are for $n = 5$. For consistency with other figures, the approximate KL Entropy is calculated for small systems of 2000 groups (shown in blue), and the exact KL Entropy is calculated for large systems of 16000 groups (shown in red). The trend lines are power laws that optimally fit the data; importantly, the scaling behavior is given by the label in the vicinity of the corresponding data (it is obviously color coded). In all these cases, the coefficients of determination are almost 1.00. Here, the markers are made especially small so that the error bars are relatively noticeable.

Regarding the standard errors, we again notice that a higher value of the KL Entropy typically has a larger error bar; we indeed mention this in the context of Figure 4, and this was also an observation in ref 38. It is also quite apparent in both Figures 10 and 11 that the error bars are somewhat negligible for the approximate KL Entropy and quite noticeable for the exact KL Entropy: Reference 38 exemplifies that this is a typical signature for the standard errors between the approximate and exact calculations, since the latter involves averaging over an exponential function, which always involves considerable fluctuations.

Regarding Figure 10, we observe very clear temperature trends for the KL Entropy. The approximate KL Entropy scales with $\beta^{2.0}$ for the monomers and with $\beta^{1.9}$ for the dimers. The scaling for the exact KL Entropy varies between $\beta^{1.14}$ and $\beta^{1.31}$ across the various alkanes. The fact that $š_{KL}$ scales almost quadratically with inverse temperature is of no surprise considering the approximate formula of eq 13 (the variance of $\Delta \tilde{U}$ appears relatively constant with $\beta$ for all of our systems). The exact formula of eq 12 formally retrieves this expression just for the limit of $\beta \to 0$, and since our study is for finite temperatures, $s_{KL}$ for our alkanes does not achieve such a quadratic scaling (although it is still superlinear). Interestingly, in multiscale simulations of the Ising model (mapped as a mean-field), this quadratic trend was formally observed at the limit of infinite temperature for both the exact and approximate variants of the KL Entopy[37]. In our current study of alkanes in the vicinity of ambient conditions, we just find the quadratic relationship for $š_{KL}$, while $s_{KL}$ just exhibits a slightly superlinear trend.

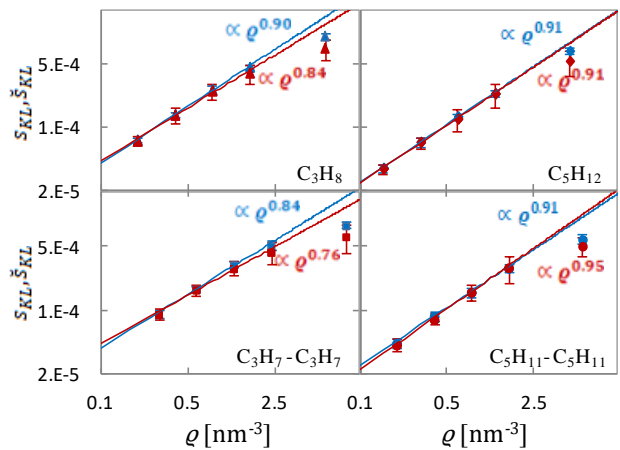

Figure 11. The KL Entropy as a function of the group density. All the formatting here is equivalent with Figure 10. Note that in fitting these power laws, the data for the highest group densities are always omitted. In all these cases, the coefficients of determination are greater than 0.98.

Regarding Figure 11, we also observe clear density trends for the KL Entropy if $\varrho \to 0$. For low densities, the values of the approximate and exact KL Entropies are almost identical, which is aligned with our previous observation that for $s_{KL} < 5 \cdot 10^{-4}$, $š_{KL} \cong s_{KL}$. Indeed, for purposes of fitting the power laws, we excluded the standard density (the highest density), since it does not correspond with the same scaling that stems at the infinitesimal limit. As such for all alkanes, we got that the approximate KL Entropy scales roughly with $\varrho^{0.9}$. The exact KL Entropy has a scaling between $\varrho^{0.76}$ and $\varrho^{0.95}$. Noteworthy, for a study of gases that deviate off the ideality assumption, it was analytically shown that the KL Entropy increases linearly with density[37]. Our study of alkanes here confirms that the KL Entropy has an almost linear relationship with group density for $\varrho \to 0$, at least at an elevated temperature. As mentioned earlier, Figure 11 does not correspond with the standard temperature of 290K (it is instead for 1160K). We also performed an analogous analysis at the standard temperature, and we found no clear dependency of the KL Entropy on the group density (in fact, most of our systems were not even stable, exhibiting a liquid droplet surrounded by a gas medium). We thus suspect that the linear relationship does not hold for alkanes in the vicinity of ambient conditions.

In summary of this entire analysis in state space, the approximate KL Entropy scales almost quadratically with inverse temperature and almost linearly with group density. For the exact KL Entropy, this trend is not as obviously evident, most likely because the temperature is not infinite, and the density is not zero. Overall, this is aligned with the observations of a previous study[37].

Finally, now that we have data for other state points as well, we examine the overall relationship between the approximate and exact KL Entropies that we alluded to earlier in eq 18. Figure 12 has the approximate value of small systems on the abscissa and the exact value of large systems on the ordinate. In some way, Figure 12 can be viewed as an inversion of Figure 5, which concurrently incorporates the data of many state points. Notably here, the different colors represent different state conditions (high temperatures, low pressures, etc.), and as usual, the different shapes of the markers represent different alkanes. The black dotted line is the identity function (i.e., based on the first-order Taylor approximation), while most importantly, the black solid line is eq 18 with $\lambda^* = 640$ (i.e., based on the second-order Taylor approximation); we obtained this value for $\lambda^*$ by minimizing the sum of the squares of the errors of their logarithms for all KL Entropies. Note that $\lambda^*$ is quite similar with $\lambda_\infty$ of Figure 5: This is because the approximate $\check{s}_{KL}$ is essentially constant between $\mathcal{N} = 2000$ and the infinite limit, while the exact $s_{KL}$ almost retrieves the infinite limit at $\mathcal{N} = 16000$. Just as we mentioned in the context of Figure 5, the linear approximation is good for values that are less than $5 \cdot 10^{-4}$ ($s_{KL} \cong \check{s}_{KL}$), while the quadratic approximation (eq 18) is good for the entire range that we studied (the coefficient of determination is more than 0.99).

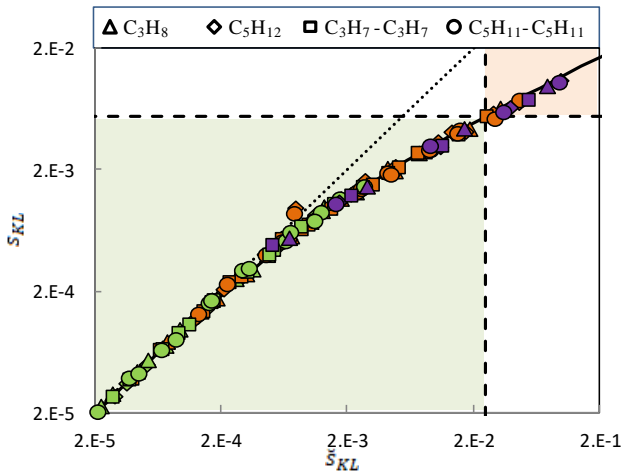

Figure 12. The exact KL entropy of large systems (ordinate) as a function of the approximate KL Entropy of small systems (abscissa). Consistent with other figures, the shape of the marker indicates the type of the alkane. The purple markers are for systems at regular conditions ($T = 290$K and $P = 15000$ Torr); the orange markers are for higher temperatures, and the green markers are for lower densities. The solid black line is eq 18 with $\lambda^* = 640$, and the dotted black line is the identity function. The green shadow rectangle indicates "good" RelRes, and the orange shadow rectangle indicates "bad" RelRes; the boundaries here (marked by dashed lines) are the same as in other figures, $\check{s}^*_{KL} = 0.0250$ and $s^*_{KL} = 0.0055$. No error bars are shown here as they are the same as those given in Figure 4, as well as in Figures 10 and 11; as the markers here are made rather large, they essentially encapsulate their corresponding error bars.

Figure 12 indeed has significant ramifications, considering how well eq 18 captures, for many state points, the relationship between $s_{KL}(\mathcal{N} = 16000)$ and $\check{s}_{KL}(\mathcal{N} = 2000)$, (i.e., the exact KL Entropy for an "infinite" system and the approximate KL Entropy for an "infinitesimal" system, respectively). On a fundamental level, the exact KL Entropy is the metric that signals multiscale errors. As we already mentioned, the exact KL Entropy is not too convenient for calculation: Besides eq 12 being prone for significant statistical uncertainties, the exact value is very much dependent on the system size; as observed in Figure 4, one must perform molecular simulations of a very large system for attaining an estimate of the "infinite" limit of $s_{KL}$. Conversely, the approximate KL Entropy is especially convenient for calculation: Equation 13 just involves the computation of an energetic variance, and the approximate value is essentially constant with system size; as observed in Figure 4, one can perform molecular simulations of a fairly small system for attaining an estimate of the "infinite" limit of $s_{KL}$.

## 4 Conclusion

In this article, we continued the work on RelRes, the multiscale approach, which combines the FG and CG models in one system by switching between their pair potentials based on the relative separation between the molecules[31,32]. Specifically here, we analyzed RelRes via the KL Entropy, which is an informatic measure for multiscale errors[37,38]. Employing the recent implementation of RelRes in LAMMPS[33], we studied the same basic alkanes that we examined previously. For each alkane, we ultimately ask: which is the optimal value of the switching distance between the FG and CG potentials? In this work, we specifically aim at answering this via the KL Entropy. Here, we examined switching distances from 0.5nm to 0.8nm, and we ultimately performed this analysis for many (fairly high) temperatures and many (fairly low) densities.

Foremost, we investigated the dependency of the (intensive) KL Entropy on the system size in Figure 4. While its exact value, $s_{KL}$, (computed via eq 12) is clearly a function of system size, we interestingly found that its approximate value, $š_{KL}$, (computed via eq 13) is practically a constant with system size. Of course, we are ultimately interested in the KL Entropy for $\mathcal{N} \to \infty$: As such, most of our current work focuses on calculating the approximate value for an "infinitesimal" system ($\mathcal{N} = 2000$) and the exact value for an "infinite" system ($\mathcal{N} = 16000$). Realize that the latter computation is especially burdensome, since eq 13 deals with averaging over exponentials, which always involves much statistical uncertainty (eq 12 clearly does not have such issues).

Predicting the exact "infinite" KL Entropy via the computation of the approximate "infinitesimal" KL Entropy becomes a main aim of this current work, and for such purposes, we derive the Taylor-based expression of eq 18 (it is the inverse of eq 17). While the fitting parameter $\lambda_{\mathcal{N}}$ appears here (this is for relating the exact and approximate KL Entropies of the same size $\mathcal{N}$), we also concurrently define the idealized parameter $\lambda^*$ for connecting the exact value of the KL Entropy for $\mathcal{N} = 16000$ and the approximate value of the KL Entropy for $\mathcal{N} = 2000$. We notably validate this empirical relation in Figure 12, which is one of the most important findings here: Across all temperatures and densities that we examined here, eq 18, with the optimal coefficient of $\lambda^* = 640$, successfully predicts $s_{KL}$ via $š_{KL}$, for all of our studied alkanes, with all of their switching distances. We reiterate that this empirical relationship means that one may never need calculate the exact KL Entropy for a very large system (with many data samples), since one can just calculate the approximate KL Entropy for a very small system (with few data samples), and by employing eq 18 (with $\lambda^* = 640$), one can obtain an excellent estimate for the fundamental metric of multiscale errors as defined by eq 10.

The main purpose of computing the KL Entropy is in holistically capturing the various multiscale errors (in structural and thermal behavior) that are associated with the RelRes system, in comparison with the reference system. For the alkanes in our study, we indeed show that the various multiscale errors more or less follow a monotonic relationship with the KL Entropy. We in turn establish bounds for the KL Entropy that ensure sufficient representation of the reference system by the RelRes system: $š_{KL}^* = 0.0250$ and $s_{KL}^* = 0.0055$. In almost all of our cases (as shown in Figures 8-9), these upper boundaries yield less than 5 percent error in the arithmetic mean and the standard deviation of the pairwise energy, as well as its associated virial (defined by eqs 15 and 16); also in all of our cases (as shown in Figures 7), these upper boundaries guarantee a value less than $5 \cdot 10^{-4}$ for the functional $\mathcal{s}_{KL}^*$ (defined by eq 14), which measures the discrepancy in the radial distribution. Consequently, based on these bounds that we suggest here, the ideal $r_s$ for a RelRes system can be obtained: Given an arbitrary molecule (governed by the LJ potential), one needs calculate the KL Entropy for several switching distances, and the ideal value of $r_s$ is the one that roughly corresponds with $\check{s}_{KL}^* = 0.0250$ and $s_{KL}^* = 0.0055$. Of course, if other researchers are interested in increasing or decreasing the magnitude of the error, which we set as our standard here, that is surely feasible: Based on our Figures 7, as well as Figures 8 and 9, one can just pick the boundaries for the KL Entropy as personally desired, and this in turn allows for the corresponding choice of the switching distance.

Confirming our previous study of alkanes[33], we also find here that using a switching distance of ~ 0.6-0.7 nm in RelRes retrieves the behavior of the conventional approach with excellent certainty. In ref 33, such $r_s$ speeds up molecular simulations by a factor of ~ 4-5, given that the cutting distance is 1.3 nm. The RelRes algorithm was especially developed for non-uniform fluids, which require higher $r_c$[31]: It can be typically ~ 1.5-1.6 nm for multi-component systems[45-48] and ~ 2.0-2.3 nm for multi-phase systems[49,50]; in such cases, RelRes will have an even greater improvement on the computational efficiency. Of course, the benefit for the computational efficiency is lesser for uniform (single-component and single-phase) systems, which employ lower $r_c$, like ~ 0.9-1.2 nm[51-58]. In summary, one can expect for nonpolar systems that RelRes will yield a gain in computational efficiency of almost an order of magnitude, while keeping an adequate representation of the static and dynamic behavior of a system of interest.

Besides, we suspect that the universality of the KL Entropy as a metric for multiscale errors suggests that it can also be utilized for studying other molecules, most notably polar ones with charges[59,60]. As mentioned earlier, Izvekov and Voth employed their variant of RelRes on such molecules that use the Coulombic potential together with the LJ potential: They notably parametrized between the FG and CG models by a numerical optimization that is based on "force matching"[29]. Regarding our formulation of RelRes, it has already been formalized theoretically for the Coulombic potential (although it has not yet been applied on any polar systems): Just as with the LJ potential, the RelRes parametrization for the Coulombic potential is also completely analytical, based on the multipole series, and thus, it does not necessitate a computer cluster while preparing for the multiscale procedure[32]. So how does the analytical parametrization roughly look like for the Coulombic potential? Most importantly, eq 3 (which forms the basis for eqs 6 and 7 of the LJ potential) must be replaced with the multipole formula presented in ref 32 (in that paper, it is given by eqs 14 and 15). For an ionic molecule (e.g., nitrate, ammonium, etc.), we expect that the zero-order term in the Taylor series is still sufficient, and in such a case, we just have Coulombic analogs for eqs 6 and 7. However, for a neutral molecule (e.g., water), an elaborate approach is necessary because its monopole vanishes: The first-order and second-order terms in the Taylor series must be incorporated, and in such a case, the Coulombic extensions for eqs 6 and 7 will have directional components that emerge through dipoles, quadrupoles, etc.; we suspect that by having such anisotropic potentials, RelRes can overcome transferability and representability issues for polar fluids, just as it did for nonpolar ones. This is a practical route for future research.

## Acknowledgements

We express our deep gratitude for the comprehensive comments of William Noid, who generously reviewed our manuscript.